\documentclass[aps,nofootinbib,showpacs,showkeys,preprint
] {revtex4}

\usepackage{epsf,epsfig,subfigure,graphicx,amsmath,amssymb}
\usepackage{color}

\def\lsim{\mathrel{\rlap{\lower4pt\hbox{\hskip1pt$\sim$}}
    \raise1pt\hbox{$<$}}}         
\def\gsim{\mathrel{\rlap{\lower4pt\hbox{\hskip1pt$\sim$}}
    \raise1pt\hbox{$>$}}}

\newcommand{\be}{\begin{eqnarray}}
\newcommand{\ee}{\end{eqnarray}}
\newcommand{\bea}{\begin{eqnarray}}
\newcommand{\eea}{\end{eqnarray}}

\begin{document}



\title{Radiatively decaying scalar dark matter through U(1) mixings
\\and the Fermi $130$ GeV gamma-ray line}

\author{Jong-Chul Park${}^a $\footnote{jcpark@kias.re.kr}
and Seong Chan Park${}^b $\footnote{s.park@skku.edu}}

\address{${}^a$ Korea Institute for Advanced Study, Seoul 130-722, Korea
\\
${}^b$ Department of Physics, Sungkyunkwan University, Suwon 440-746, Korea}

\begin{abstract}
In light of the recent observation of the Fermi-LAT 130 GeV
gamma-ray line, we suggest a model of scalar dark matter in a
hidden sector, which can decay into two (hidden) photons. The
process is radiatively induced by a GUT scale fermion in the loop,
which is charged under a hidden sector U(1), and the kinetic
mixing ($\sim \epsilon F^{\mu\nu}F'_{\mu\nu}$) enables us to fit
the required decay width for the Fermi-LAT peak. The model does
not allow any dangerous decay channels  into  light standard model
particles.
\end{abstract}




\maketitle


\section{Introduction}

Although the major constituent of matter content of the Universe
is dark matter (DM), we  only  know little about its nature so
far~\cite{Jungman:1995df}. Having no DM candidate in its particle
contents, the standard model (SM) is strongly required to be
extended. One intriguing possibility is that a hidden sector
attached to the standard model sector  includes a dark matter
candidate. The singlet DM candidate could be a
scalar~\cite{Scalar}, a fermion~\cite{Fermion, millicharged,
511millicharged} or a vector boson~\cite{Vector, Chen}.

The recent claim of the Fermi Large Area Telescope
(Fermi-LAT)~\cite{Fermi} 130 GeV $\gamma$-ray line
~\cite{Weniger:2012tx, Bringmann:2012vr} shed light on further
details of the dark matter property since no known astrophysical
source would produce such a peak. The claim was  further
strengthened by~\cite{Su:2012ft} (and also~\cite{Tempel:2012ey,
Tempel2}) even though the Fermi-LAT collaboration only has
provided sensitivity limits on dark matter models based on a part
of the acquired data set on different region of
interest~\cite{Ackermann:2012qk} (also see
\cite{GeringerSameth:2012sr}). There are explanations of this
$\gamma$-ray line based on spectral and spatial variations of
diffuse $\gamma$-ray~\cite{Boyarsky:2012ca} and new background
with `Fermi-bubble'~\cite{FermiBubble}, but the most interesting
interpretation might be that the $\gamma$-ray line could be
originated from the DM annihilation, $\chi \chi \to \gamma\gamma\,
[{\rm or}\, \gamma X]$ with $E_\gamma \simeq m_\chi\, [{\rm or}\,
m_\chi(1-m_X^2/4m_\chi^2)] \simeq 130$ GeV~\cite{Ibarra:2012dw,
Dudas:2012pb, Cline:2012nw, KYChoi, Lee:2012bq, Rajaraman:2012db,
Acharya:2012dz, Buckley:2012ws, Chu:2012qy, Das:2012ys,
Kang:2012bq, Oda:2012fy, Yang:2012ha} or the DM decay, $\chi \to
\gamma \gamma\, [{\rm or}\, \gamma X]$ with $2E_\gamma \simeq
m_\chi [{\rm or}\, m_\chi(1-m_X^2/m_\chi^2)] \simeq 260$
GeV~\cite{Kyae:2012vi, Kang:2012bq}.
\footnote{If the Fermi-LAT peak is from the ``box-shaped" spectrum as was discussed in \cite{Ibarra:2012dw}, the energy of gamma-ray $E_\gamma$ would not be $m_\chi$ or $m_\chi/2$. In
Refs.~\cite{Buchmuller:2012rc, Cohen:2012me, Cholis:2012fb}, it
has been studied to constrain the DM models accompanied by
continuum photons correlated to the 130 GeV gamma-ray line.}

For the DM annihilation interpretation of the 130 GeV $\gamma$-ray
peak, the required values for annihilation cross section is found
to be $\langle\sigma
v\rangle_{\chi\chi\rightarrow\gamma\gamma}\sim a few \times
10^{-27} ~{\rm cm^3/s}$,\footnote{More precisely, $\langle\sigma
v\rangle_{\chi\chi\rightarrow\gamma\gamma}\simeq  (1.27\pm
0.32^{+0.18}_{-0.28})\times 10^{-27}~ {\rm cm^3/s}$ ($2.27\pm
0.57^{+0.32}_{-0.51}\times 10^{-27}~ {\rm cm^3/s}$)  for the
Einasto \cite{Einasto} (NFW \cite{NFW}) DM profile~\cite{Weniger:2012tx}.} which is
approximately one order of magnitude smaller than the total
annihilation cross section for the thermal production of DM,
$\langle \sigma v\rangle_{\chi\chi\to SM}\simeq 3\times
10^{-26}~{\rm cm^3/s}$~\cite{Jungman:1995df}. As the dark matter
is likely to be electrically neutral (or
milli-charged~\cite{millicharged, 511millicharged,
Goldberg:1986nk}), the annihilation process for $\gamma\gamma$
production may be radiatively induced by massive charged particles
in the loop. If some of the charged particles are lighter than the
DM particle, there could appear tree-level annihilation channels
to these charged particles, which may dominantly determine the
relic abundance of dark matter. However, the loop factor is too small as
$g^2/16\pi^2 \lsim 10^{-2}$ and thus does not correctly account the
discrepancy between the cross sections. A variety of annihilating
DM models have been suggested to overcome this
issue~\cite{Ibarra:2012dw, Dudas:2012pb, Cline:2012nw, KYChoi,
Lee:2012bq, Rajaraman:2012db, Acharya:2012dz, Buckley:2012ws,
Chu:2012qy, Das:2012ys, Kang:2012bq, Oda:2012fy, Yang:2012ha}.

Decaying DM can be an alternative explanation. Indeed, decaying
dark matter models have been  recently proposed to account the
excessive observation of positron in the PAMELA and ATIC where the
dark matter is a vector boson in a hidden sector~\cite{Chen}. The
vector boson of the hidden sector Abelian gauge group ${\rm
U(1)'}$ can decay to the standard model photon through the
kinematical mixing term $\epsilon F_{\mu\nu}F'^{\mu\nu}$, where
$F^{\mu\nu}(F'^{\mu\nu})$ is the field strength tensor of ${\rm
U(1) (U(1)')}$ gauge boson, respectively. As the mixing parameter
could be small ($\epsilon \sim 10^{-26}$) ~\cite{Chen}, the decay
width could be suppressed. However, the decay of a vector boson to
a pair of photons is forbidden by the Landau-Yang
theorem~\cite{LandauYang} so that we need another model. In Refs.
~\cite{Kyae:2012vi, Kang:2012bq}, a scalar dark matter, $\phi$,
was considered with an effective operator allowing the decay to
two photons: $c_6\frac{\phi \phi}{\Lambda^2}
F^{\mu\nu}F_{\mu\nu}$, which is dimension six.\footnote{For
radiative DM decays to $\gamma\gamma$ or $\gamma X$, see
e.g.~\cite{Garny:2010eg}.} It is pointed out in Ref.
\cite{Kyae:2012vi} that a dimension five operator, $c_5
\frac{\phi}{\Lambda} F^{\mu\nu}F_{\mu\nu}$,  cannot fit the data
without introducing Trans-Planckian cutoff ($\Lambda \gg M_{Pl}$)
or equivalently a largely suppressed coefficient $c_6\ll 1$ as the
required partial decay width of the dark matter to photons is
extremely small, $\Gamma(\phi \to \gamma \gamma)\sim 10^{-29}
s^{-1}$.

In this paper, we try to combine the advantages of above two cases:
\begin{itemize}
\item A scalar dark matter {\it can} decay into $\gamma\gamma$
    differently from the massive vector dark matter,
\item A small kinetic mixing $\epsilon$ can make the
    effective couplings of the dark matter particle with the
    standard model particles small.
\end{itemize}
Combining these two advantages, we suggest a dark matter model,
which has an effective dimension five operators of the form:
\begin{eqnarray}
{\cal O} = c_5 \frac{\phi}{\Lambda}  \left(  F'_{\mu\nu}F'^{\mu\nu}, +  \epsilon F_{\mu\nu}F'^{\mu\nu} + \epsilon^2 F_{\mu\nu}F^{\mu\nu}\right),
\end{eqnarray}
from which we can learn that the decay amplitudes to
$\gamma\gamma'$ and $\gamma\gamma$ are relatively suppressed by a
factor of $\epsilon$ and $\epsilon^2$ with respect to the one for
$\gamma'\gamma'$ channel due to the mixing.

In the next section (Sec. \ref{sec:bounds}), we further explain
the model in detail and present the partial decay widths of the
dark matter to (hidden) photons then clarify the model parameter
space providing a good fit  to the 130 GeV gamma-ray line.
Discussions on possible experimental bounds on the same parameter
space follow. In Sec. \ref{sec:issues}, we further discuss the
theoretical issues concerning the consistency of the model and
also other cosmological observations then conclude in Sec.
\ref{sec:conclusion}. Finally, in Appendix, we present some
details of the U(1) mixing Lagrangian for an extra unbroken U(1)
symmetry.

\section{The model and experimental bounds}
\label{sec:bounds}

Postulating extra $\rm U(1)$ gauge symmetries is one of the
simplest extensions of the standard model. As the kinetic mixing
term $\epsilon F^{\mu \nu} F'_{\mu\nu}$ is compatible with Lorentz
as well as gauge symmetry, the term should be included in view of
effective field theory. The term can be generated through one-loop
diagrams with a bi-charged fermion~\cite{KineticMixing}. If the
extra U(1) is broken by a hidden sector Higgs mechanism, the gauge
boson ($Z'$) gets mass and mixes with the standard model $Z$ boson
\cite{Langacker08}. A general analysis for a hidden sector DM,
which is charged under a broken $\rm U(1)_H$ was done in
Ref.~\cite{Chun:2010ve}. On the other hand, if the extra U(1) is
exact, the massless gauge boson (hidden photon or exphoton,
$\gamma_H$) can mix with the usual photon ($\gamma$) and the
corresponding phenomenology becomes quite different from the case
with $Z'$. It is worth noticing that a light DM explanation with
the hidden photon  for the anomalous 511 keV $\gamma$-ray
signature from the Galactic Center was considered in
Ref.~\cite{511millicharged}, where a milli-charged fermion dark
matter was introduced.

For the Fermi-LAT $\gamma$-ray peak, a scalar DM is more
profitable. The minimal setup only introduces a heavy vector-like
fermion $\psi$, which is only charged under the $\rm U(1)_H$, and
a scalar dark matter candidate $\phi$.  The hidden sector can
interact with the SM sector through a kinetic mixing between $\rm
U(1)_{EM}$ and $\rm U(1)_H$.\footnote{Here the `Higgs-portal'
interaction, $\phi^2 |H|^2$, is assumed to be negligible as we do
not allow the DM decay via mixing with the Higgs boson. Further
discussion is given in Sec. \ref{sec:issues}. } The model
Lagrangian is given by
\begin{eqnarray}\label{Interactions}
{\cal L} \supset {\cal L}_{SM} - \frac{1}{4}\, \hat{F}_{H \mu\nu}\hat{F}_H^{\mu\nu}
- \frac{\sin\epsilon}{2}\,  \hat{F}_{\mu\nu}\hat{F}^{\mu\nu}_H - \lambda \phi \overline{\psi}\psi + i
\overline{\psi}\gamma^\mu (\partial_\mu - i \hat{g}_H \hat{A}^H_\mu) \psi -
m_\psi \overline{\psi}\psi\,,
\end{eqnarray}
where $\hat{F}^{\mu\nu}$ and $\hat{F}^{\mu\nu}_H$ are respectively
field strength tensors for $\rm U(1)_{EM}$ and $\rm U(1)_H$. The
detailed derivation of the kinetic mixing between $\rm U(1)_{EM}$
and massless $\rm U(1)_H$ can be found in Appendix~\ref{Appendix}.
Current bounds on hidden $\rm U(1)$ gauge bosons and milli-charged
particles (MCPs) are well summarized in Ref.~\cite{HiddenPhoton}.
Indeed, when $m_{\gamma_H} = 0$, the only change for the SM photon
interactions is the modification of the coupling constant,
$\hat{e} \rightarrow \hat{e}/\cos \epsilon$, which can be simply
refined by a field redefinition.

\begin{figure}[t]
\begin{center}
\includegraphics[width=0.32\linewidth]{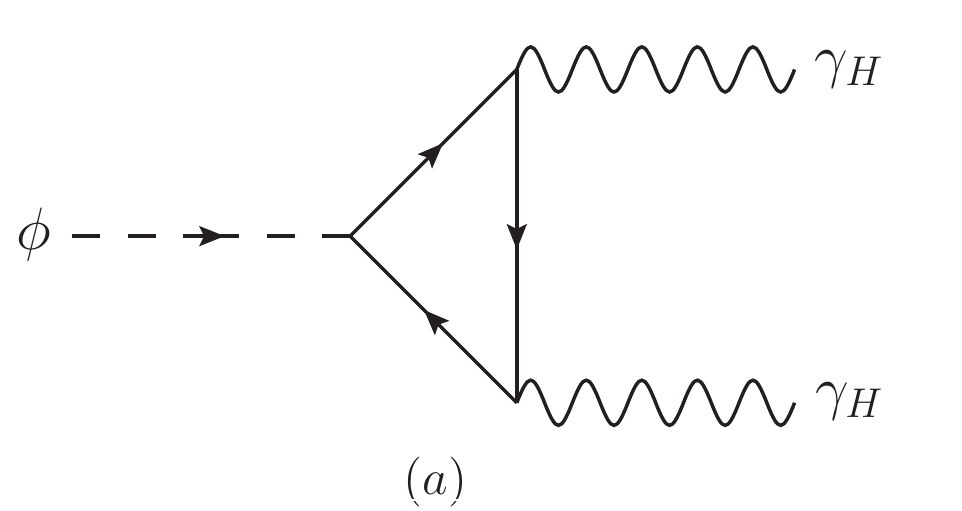}
\includegraphics[width=0.32\linewidth]{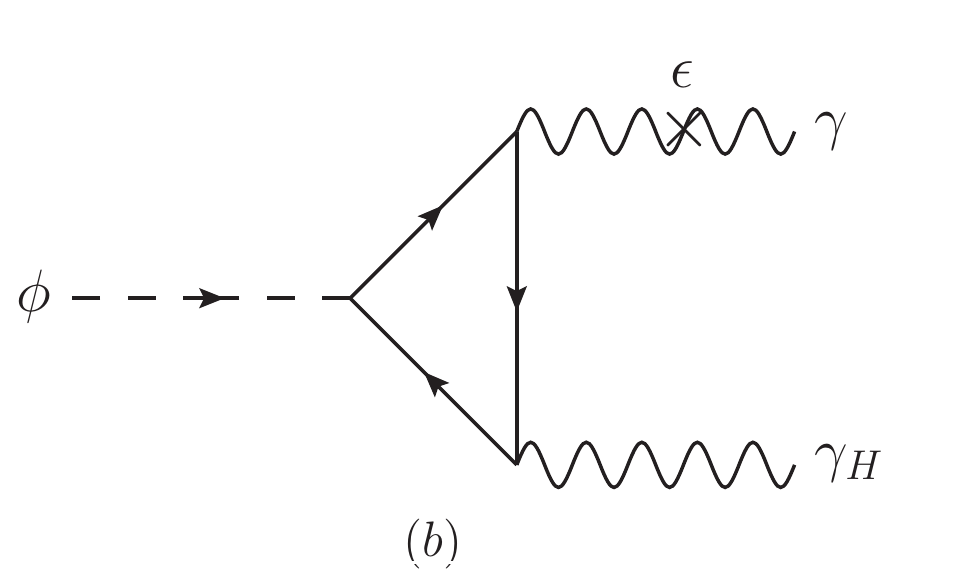}
\includegraphics[width=0.32\linewidth]{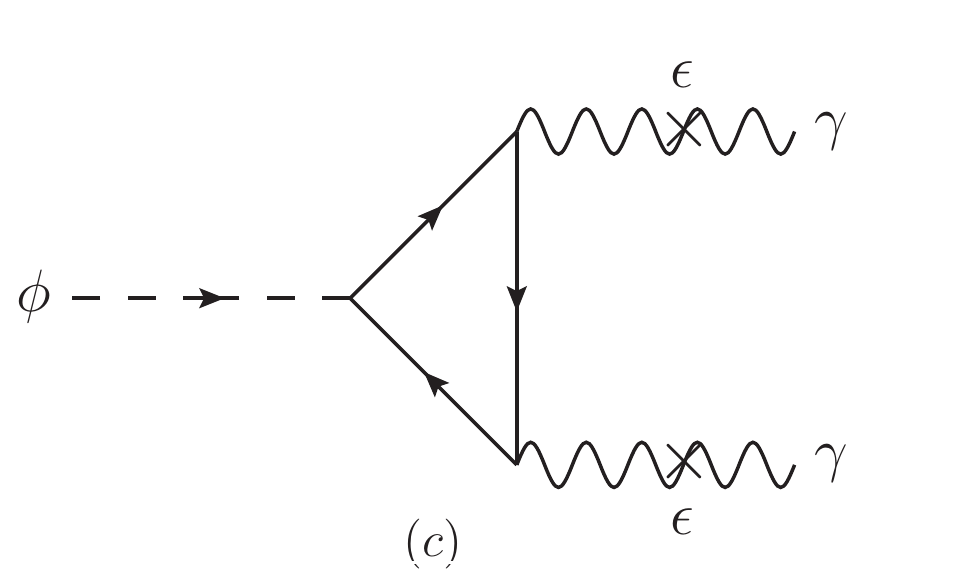}
\end{center}
\caption{(a) Scalar dark matter decaying to two hidden photons.
(b) Scalar dark matter decaying to one hidden photon and one SM photon.
(c) Scalar dark matter decaying to two SM photons. }
\label{Fig_DMdecay}
\end{figure}

The scalar DM candidate $\phi$ can radiatively decay to the hidden
photon ($A_H = \gamma_H$) as well as the conventional photon
($\gamma$) of the EM interaction through the triangle diagrams
with the virtual hidden sector fermion $\psi$ as can be seen from
Fig.~\ref{Fig_DMdecay}. From the couplings in
Eq.~(\ref{Interactions}), one can easily calculate the decay width
of $\phi$:
\begin{eqnarray}
\Gamma(\phi \to \gamma_H \gamma_H) = \frac{\left(\alpha_H
\lambda\right)^2}{256\pi^3}\frac{m_\phi^3}{m_\psi^2}|F(\tau)|^2\,,
\end{eqnarray}
where $\alpha_H = g_H^2/4\pi$ and $F(\tau= 4m_\psi^2/m_\phi^2) =
-2 \tau \left[1 + (1 - \tau)\arcsin^2(1/\sqrt{\tau})\right]$ which
is well approximated by $-4/3$ at a large $\tau$ limit.
To make $\phi$ stable enough, we would require the longevity of
$\phi$:
\begin{eqnarray}\label{eq:gamma}
\Gamma(\phi \to \gamma_H \gamma_H)^{-1} \gg \tau_{\rm Universe}
\approx 4.34 \times 10^{17}\, {\rm sec}\,.
\end{eqnarray}
Thus, we can obtain a constraint on
the combination $\alpha_H \lambda$:
\begin{eqnarray}
\alpha_H \lambda \ll 1.96 \times 10^{-7} \left( \frac{m_\psi}{10^{16}\, {\rm GeV}} \right)
\quad {\rm for} \quad m_\phi = 260 {\rm GeV}\,.
\end{eqnarray}
%

%
%

The DM candidate $\phi$ predominantly decays to two hidden
photons, $\phi \to \gamma_H \gamma_H$
[Fig.~\ref{Fig_DMdecay}-(a)], which thus determines the life time
of $\phi$ but  a hidden photon can be converted into a SM photon
through the kinetic mixing. Consequently, one can detect the DM
decay signals through the decay modes $\phi \to \gamma \gamma_H/
\gamma \gamma$ [Fig.~\ref{Fig_DMdecay}-(b)/(c)], which are
respectively suppressed by $\epsilon^2$ and $\epsilon^4$ compared
with the dominant decay mode:
\begin{eqnarray}
\Gamma(\phi \to \gamma_H \gamma_H):\Gamma(\phi \to \gamma
\gamma_H) : \Gamma(\phi\to \gamma\gamma) \simeq 1:\epsilon^2
:\epsilon^4\,.
\end{eqnarray}
This helps as the effective dimension five operators, $\phi  F_{\mu\nu} F^{\mu\nu}$ and $\phi  F_{\mu\nu} X^{\mu\nu}$, are all suppressed by powers of $\epsilon$ and provides the required decay rate
$\Gamma^{-1} \approx {\cal C} \times 10^{29}$ sec~
\cite{Kyae:2012vi, Buchmuller:2012rc} or
\begin{eqnarray}
\Gamma(\phi \to \gamma_H \gamma)^{-1} \approx 1.52 {\cal C} \times 10^{53} \,{\rm GeV}^{-1}\,,
\end{eqnarray}
where a convenient parameter ${\cal C}\in (0.1,1)$ is introduced.
Then, the parameter range for the  mixing parameter $\epsilon$ can be read:
\begin{eqnarray}
\epsilon \approx \frac{4.1 \times 10^{-13}}{\alpha_H \lambda \sqrt{\cal C}} \left(
\frac{m_\psi}{10^{16}\, {\rm GeV}} \right) \quad {\rm for} \quad
m_\phi = 260 {\rm GeV}\,.
\end{eqnarray}
Having a small value for  $\alpha_H \lambda \ll 10^{-7}$, a
relatively sizable kinetic mixing parameter $\epsilon$ is
required.

%
\begin{figure}[t]
\begin{center}
\includegraphics[width=0.80\linewidth]{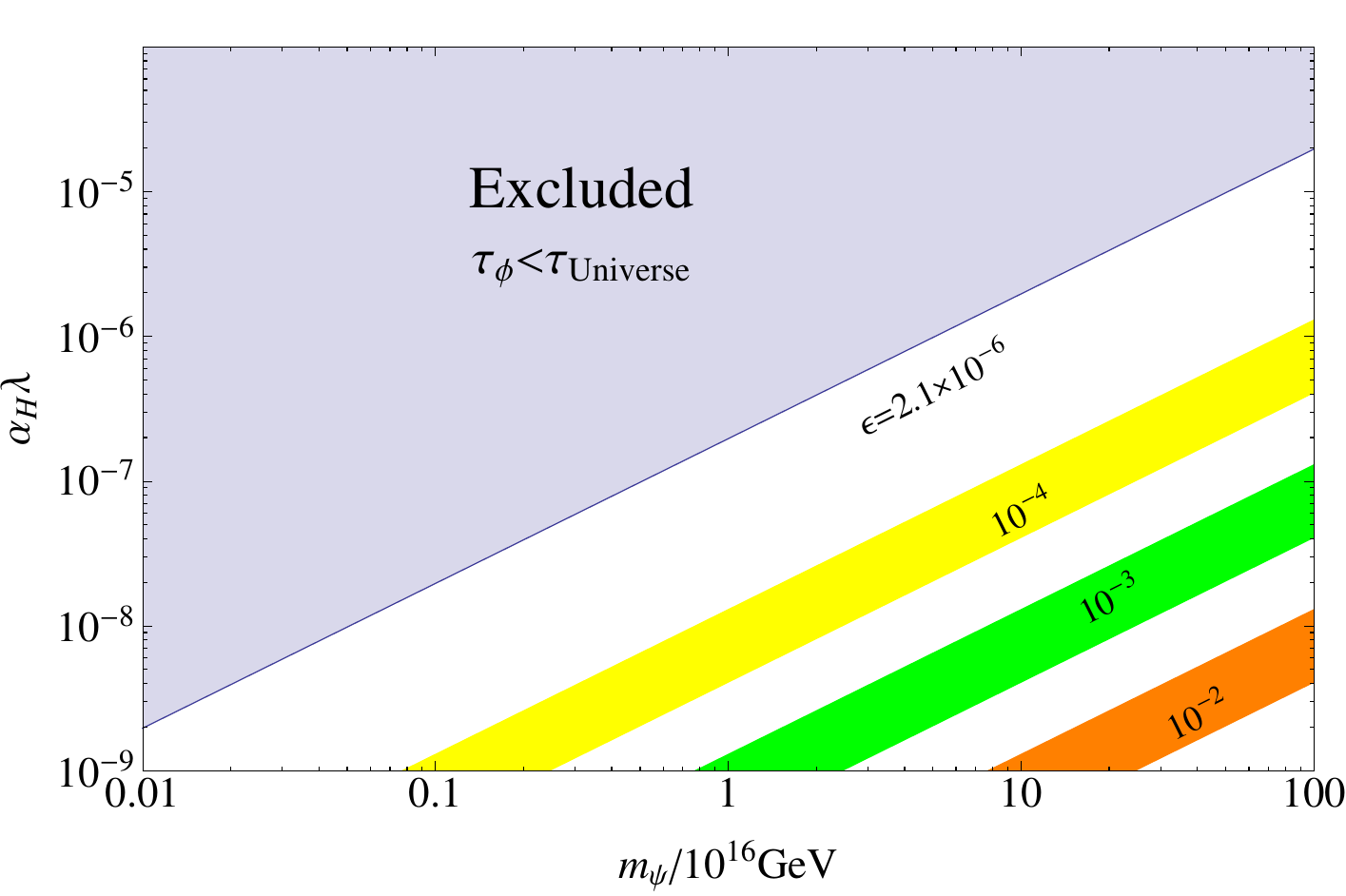}
\end{center}
\caption{Contour plot for the kinetic mixing parameter $\epsilon$
in the $m_\psi-\alpha_H \lambda$ plane. The colored region above
the thick  line is excluded since the life time of $\phi$ is
shorter than the age of the Universe. The partial life time of
$\phi$ to $\gamma \gamma_H$ is fixed to be in  $(10^{28},10^{29})
\,{\rm sec}$, which  is required to fit the Fermi-LAT data,
with $\epsilon =10^{-4}-10^{-2}$ in the colored bands from the
top to the bottom on the right lower side of the graph. The
larger $\alpha_H \lambda$ is required for the smaller $\epsilon $.
} \label{fig:plot}
\end{figure}
%

In Fig. \ref{fig:plot}, we plotted the parameter space in the
$m_\psi-\alpha_H \lambda$ plane. The upper left region in shade
(grey) is excluded by the longevity of the dark matter. The three
colored bands are for fitting the required decay width
($\Gamma^{-1}=10^{28}-10^{29} \,{\rm sec}$) with $\epsilon
=10^{-4}, 10^{-3}$ and $10^{-2}$, respectively. The charged
fermion $\psi$ is assumed to be heavy ($\sim 10^{16}$ GeV).


\section{Morphology of the 130 GeV $\gamma$-rays: Decay vs Annihilation}

Based on the information about the spatial distribution of the 130
GeV gamma-ray line, we can compare decaying dark matter with
annihilating dark matter. As the observed gamma-ray flux would
depend linearly on the dark matter density for decaying dark
matter ($\propto n_{DM}$), the characteristic morphology of the
expected gamma-ray from decaying dark matter is flatter than the
one from the annihilating dark matter, which is quadratically
sensitive to the density ($\propto n_{DM}^2$). Taking the spatial
distribution of the Fermi-LAT excess at 130 GeV into account,
which is relatively peaky in the Galactic Center, indeed the
decaying dark matter would require an enhancement of the density
near the Galactic Center, in comparison with conventional profiles
e.g., NFW and Einasto profiles \cite{Buchmuller:2012rc}. However,
there still exists large uncertainty of the dark matter density
close to the Galactic Center and large systematic and statistical
uncertainties at present. Thus, the decaying dark matter could
remain as a possible solution to the Fermi-LAT excess as the
similar conclusion was made in Ref. \cite{Buchmuller:2012rc}.

%
\begin{figure}[h]
\begin{center}
\includegraphics[width=0.66\linewidth]{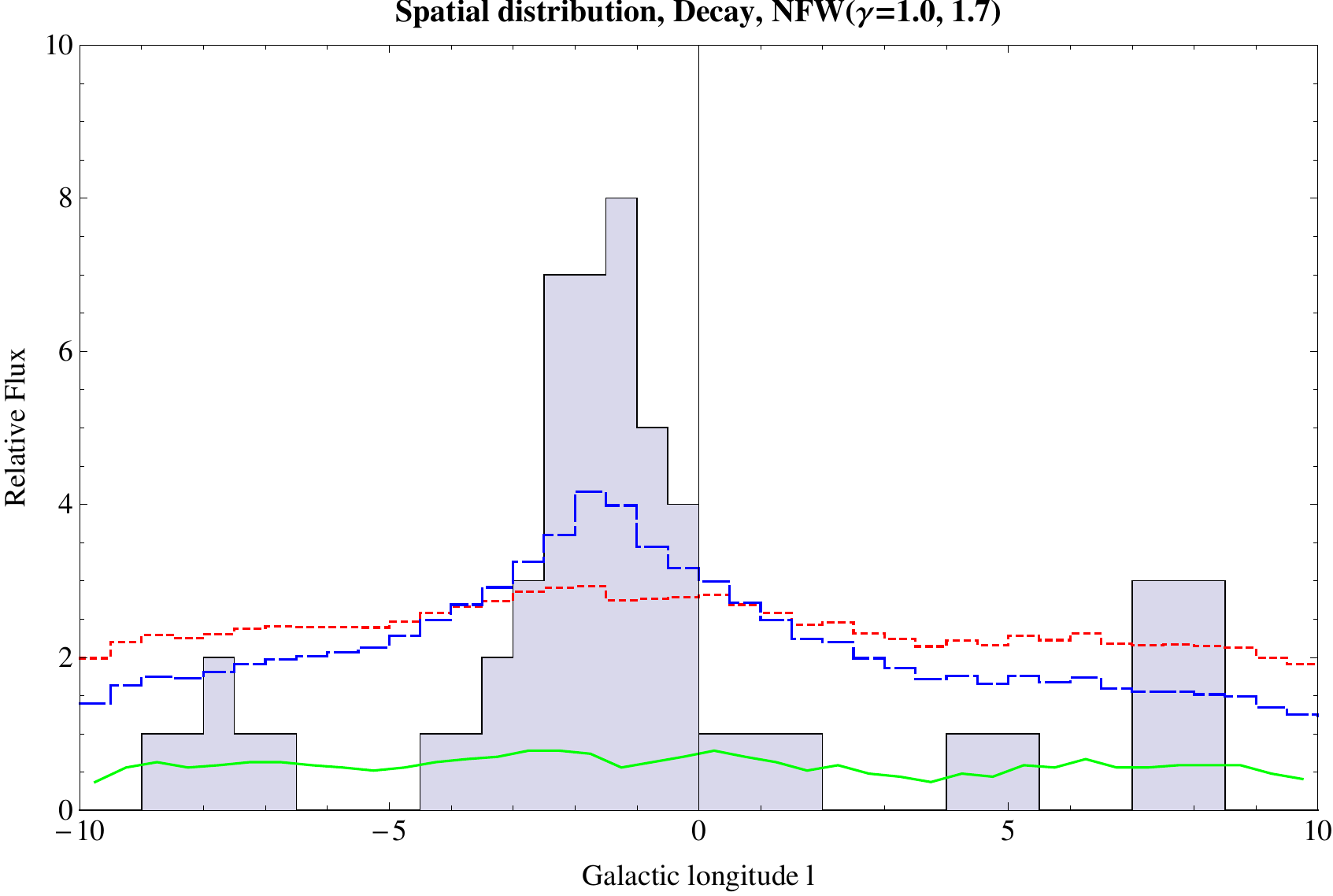}
\includegraphics[width=0.66\linewidth]{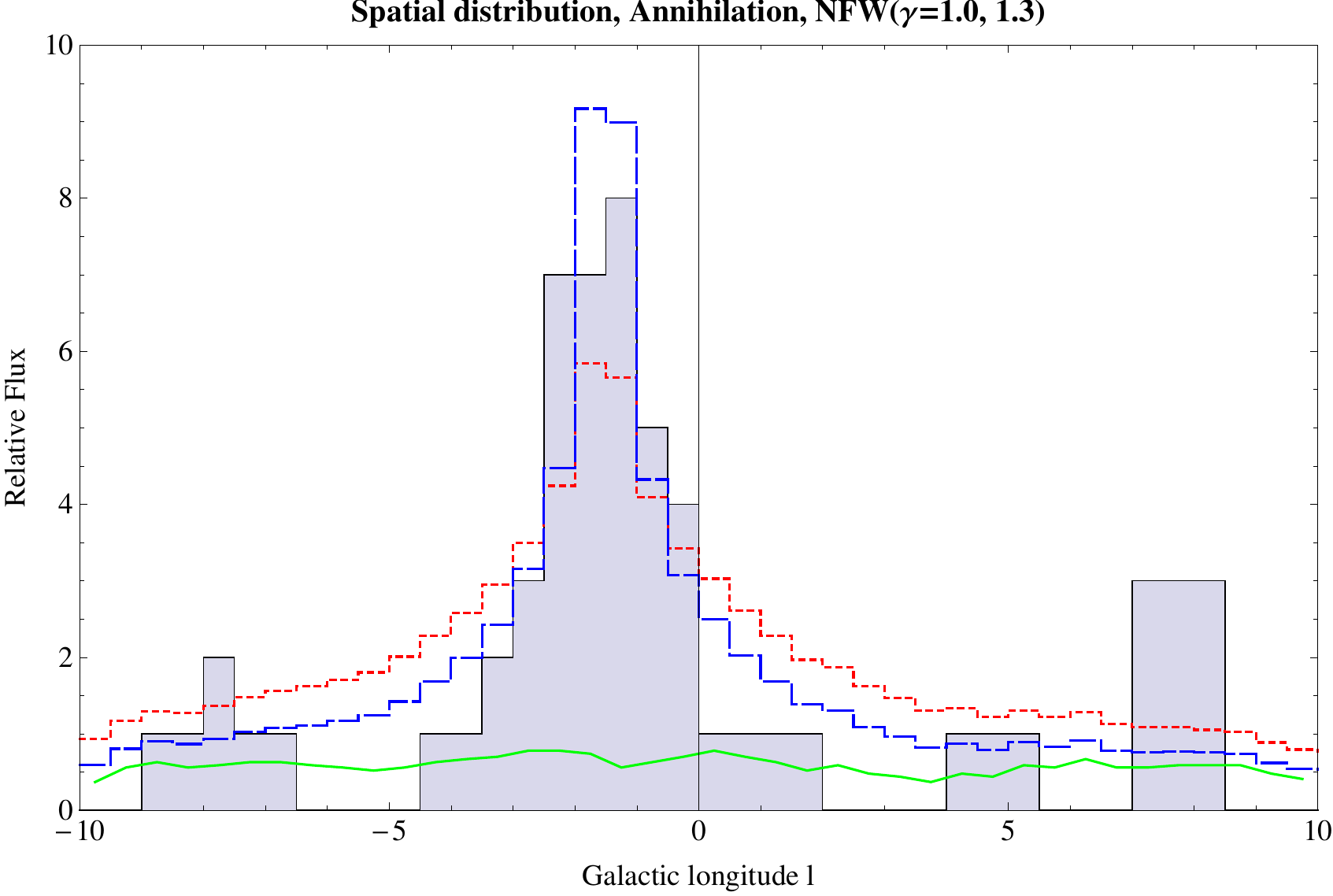}
\end{center}
\caption{The profiles of decaying (top) and annihilating (bottom) dark matter
fitting the 130 GeV gamma-ray line along the Galactic plane
($\Delta l=0.5^\circ, |b|<5^\circ$). 
Compared to the rather flat NFW profile ($\gamma=1.0$: the red dotted line), the steeper profile ($\gamma= 1.7$: the blue dashed) provides the better fit for the decaying DM. For comparison, we plotted the fits for the annihilating DM with $\gamma=1.0$ (red dotted) and $\gamma=1.3$ (blue dashed). The green line is for the background. Fermi-LAT excess data is taken from Ref. \cite{Su:2012ft}. See Table \ref{table:chi2} for $\chi^2/d.o.f.$ values for each fit.
} \label{fig:profile}
\end{figure}
%

In Fig. \ref{fig:profile}, we plotted the best fit distributions
of gamma-rays for the decaying (Top) and annihilating (Bottom) DM
within the allowed range of $\Gamma_{\rm DM}$ and $\langle \sigma v \rangle$ in Galactic longitude, $\ell \leq |10|^\circ$. 
The estimated background is depicted by green line, which is roughly $\ell$ independent. 
For fitting we used a useful parametrization of the DM profiles with
$(\alpha,\beta)=(1,3)$:
\begin{eqnarray}
\rho(r) = \rho_0 \left(\frac{r_0}{r}\right)^{\gamma} \left(\frac{1+(r_0/r_s)^\alpha}{1+(r/r_s)^\alpha}\right)^{\frac{\beta-\gamma}{\alpha}},
\end{eqnarray}
where the local density $\rho_0 \simeq 0.4 \,{\rm GeV/cm^3}$ is
given at $r_0 \simeq 8.5$ kpc. The scale radius is set to be
$r_s=20$ kpc for numerical analysis. The popular Navarro-Frenk-White (NFW) profile is obtained with $(\alpha,\beta,\gamma)=(1,3,1)$ but we allow a rather broad
parameter range $\gamma  \in [1.0, 1.7]$. With the bigger $\gamma$, the profile becomes sharper in the center. Einasto profile is also commonly used but provides less sharp distribution thus is less favored for the decaying DM scenario. As it is expected,  the sharper profile
provides the better fit for the decaying DM.  When $\gamma=1.7\,(1.5, 1.0)$,  the best fits are found with  the decay widths $\Gamma^{-1}=2.8 \,(2.1, 1.3) \times 10^{28}$ sec, respectively.

The overall quality of fit is parametrized with $\chi^2/d.o.f.$ value as shown  in Table \ref{table:chi2}.  
Taking the background effect into account, we found that the best fit is found with a rather small $\gamma$ for the annihilating DM but the larger $\gamma$ for the decaying DM. It is worth noticing that the decaying DM provides acceptable fits with $\chi^2/d.o.f <1.80$  for {\it all} values of $\gamma \in [1.0,1.7]$ even though the annihilating DM provides better fit when $\gamma\leq 1.5$. When the profile is extremely peaky with a larger value $\gamma>1.5$, the decaying DM provides better fit than the annihilating DM. The reason is simple: as it is seen in Fig. \ref{fig:profile}, the gamma-rays are found not only within $|\ell|\leq 5^\circ$ but also in rather broad range of $5^\circ-10^\circ$, which can not be properly explained by the background (green line in Fig. \ref{fig:profile}). Due to those signals from the outside of $5^\circ$, a peaky profile does not necessarily help but actually makes the fitting worse in the annihilating DM case. 

In conclusion, currently both of decaying and annihilating
dark matter models are acceptable with $\chi^2/d.o.f. < 1.8$ provided that the profile 
could be  enhanced  at the center with $\gamma>1.0$ for the decaying DM model. 
Future improvement in observation is highly required to judge the fate of each model.

\begin{center}
\begin{table}[h]
\begin{tabular*}{0.5\textwidth}{ @{\extracolsep{\fill}}  c ||  c  c  c   c  }
\hline
~~NFW $\gamma$ ~~& 1.0 & 1.3 & 1.5 &  1.7~~~ \\
  \hline
~~ Ann.~~ & 1.05 & 1.03 & 1.17 & 1.40 ~~~\\
 ~~Dec.~~ &  1.80 & 1.60 & 1.44 & 1.28 ~~~\\
  \hline
\end{tabular*}
\caption{The $\chi^2/d.o.f$ values for Annihilating and decaying dark matter for various dark matter profiles with $\gamma \in [1.0, 1.7]$.}
\label{table:chi2}
\end{table}
\end{center}

\section{Other issues}
\label{sec:issues}

In this section, we discuss possible difficulty of the
`Higgs-portal' type interaction and its way out. Then, we propose
some scenarios for producing the singlet dark matter from the
inflaton or other heavy particle decays in the early universe.

\subsection{$\sigma \phi^2 |H|^2$ or $\mu \phi |H|^2$ type interactions}

The gauge invariant renormalizable interactions between the dark
matter and the Higgs boson of the form $\sim \sigma \phi^2 |H|^2$
(``Higgs portal") and $ \mu \phi |H|^2$ are generically allowed.
If $m_\phi \geq 2 m_h$, $\phi$ can therefore decay into two Higgs
bosons with the decay rate:
\begin{eqnarray}
\Gamma(\phi \rightarrow hh) \propto \frac{1}{16\pi}\, \frac{\left(\sigma \langle \phi
\rangle + \mu \right)^2}{m_\phi}\sqrt{1-\left(\frac{2 m_h}{m_\phi} \right)^2}\,,
\end{eqnarray}
which can be significantly larger than the required decay width in
Eq.~(\ref{eq:gamma}).
This difficulty can be avoided when an additional spatial
dimension exists and all the hidden sector particles ($\phi, \psi,
\gamma_H $) are localized on the  `hidden-brane' which is
spatially distant from the `visible-brane'  on which all the
SM particles ($ \supset \{H, \gamma, leptons,
quarks\}$) reside. A mediator fermion $\psi_M$, which is charged
under the both of the Abelian gauge groups ${\rm U(1)_{EM}} \times
{\rm U(1)_H}$, could provide a sizable $\gamma-\gamma_H$ mixing.
Then, the operator $\hat{\cal O}\sim \phi^2 H H^\dagger$ is suppressed by
many loop factors (two one-loops of the $\phi\phi-\gamma_H\gamma_H$
and $HH-\gamma\gamma$ vertices, and two additional one-loops of the
$\gamma-\gamma_H$ mixings) and also by possible suppressed wave function
overlaps $\sim \psi(y_{\rm hidden})^* \psi(y_{\rm visible})$.
Similarly, the other operator $\hat{\cal O}\sim \phi H H^\dagger$
is also suppressed.

\subsection{The production of $\phi$}

\subsubsection{Model-1}

A simple mechanism for the production of $\psi$ and $\phi$ is the inflaton
decay after reheating: $\Phi \to \psi \overline{\psi}$ and $\Phi
\to \phi\phi$, where $\Phi$ is the inflaton field. As the mass of
$\psi$ is around the GUT scale, we do not expect any sizable
number of remaining $\psi$ in our patch universe, but still $\phi$
could remain to be a good DM candidate. Here the only
requirement is
\begin{eqnarray}
m_\psi > T_{\rm reheating} > m_\phi\,,
\end{eqnarray}
where a large space of reheating temperature in TeV to GUT-scale
could be allowed, in principle. With this condition, $\phi$ can be
produced but $\psi$ cannot. In this paper, we simply assume that
the reheating temperature is low enough compared with the
GUT-scale to make the number of $\psi$ small enough.

\subsubsection{Model-2}

If $m_\psi \ll T_{\rm reheating}$, large number of $\psi$ can be
produced through the reheating process. In the presence of an
additional hidden sector fermion field $\psi'$, we can consider
the decay of $\psi$, $\psi \to \psi' \phi$. Then, a simple $\rm
U(1)_H$ interaction of the form $\overline{\psi'}\psi' \gamma_H$
can induce the pair annihilation of $\psi'$, $\psi'\psi' \to
\gamma_H \gamma_H$, so that the relic abundance of $\psi'$ could
be small enough provided that
\begin{eqnarray}\label{RelicLimit}
\langle \sigma v \rangle_{\psi'\psi' \to \gamma_H\gamma_H} \sim
\frac{\pi \alpha_H^2}{m_{\psi'}^2} \gg 2 \times
10^{-9} {\rm GeV}^{-2}\,,
\end{eqnarray}
for which we assume that the mass of $\psi'$ is relatively light.
For instance, if $\alpha_H\sim 10^{-4}$, $m_{\psi'} \ll 4~{\rm
GeV}$ insures that the number of relic $\psi'$ is too small to
affect the present universe. The least constrained mass range for
the hidden sector milli-charged particle is $0.1~{\rm GeV}
\lesssim m_{\psi'} \lesssim 500~{\rm GeV}$~\cite{HiddenPhoton}.
Thus, we can easily find experimentally allowed masses for $\psi'$
satisfying Eq.~(\ref{RelicLimit}).

\section{Conclusion}
\label{sec:conclusion}

The recently reported gamma-ray excesses around 130 GeV based on
the Fermi-LAT data is difficult to explain with well known dark
matter models. Inspired by the Fermi-LAT 130 GeV line, we suggest a model with a decaying scalar dark
matter $\phi$ and a heavy hidden fermion $\psi$ charged under a
hidden gauge symmetry $\rm U(1)_H$ allowing a Yukawa interaction,
$\lambda \phi \bar{\psi}\psi$.  In this model, the hidden sector
can communicate with the SM sector through the kinetic mixing
($\epsilon F^{\mu\nu} F'_{\mu\nu}$) and the unwanted fast decay to
$\gamma\gamma$ is well suppressed by the small mixing $\epsilon^2$
after ensuring the longevity of $\phi$ by a doubly suppressed loop
factor, $\Gamma_\phi/m_\phi \sim (\alpha_H \lambda)^2
(m_\phi^2/m_\psi^2) / (256\pi^3)$.  Assuming $\alpha_H \lambda
\sim 10^{-7}$ and $\epsilon \sim 10^{-2}-10^{-4}$, we found that
the required decay rate $\Gamma(\phi \rightarrow \gamma_H \gamma)
\sim 10^{-28}\, {\rm sec}^{-1}$ for the observed $\sim 130$ GeV
gamma-ray flux is obtained with a GUT scale  mass for the fermion
$m_\psi\sim 10^{16}$ GeV.

From the morphology of the observed gamma-ray, the dark matter profile is suggested to be relatively enhanced in the Galactic Center for the decaying dark matter. It is found that the decaying dark matter could provide a good fit 
producing less additional gamma-rays  outside of the Galactic Center when $\gamma>1$, which is still compatible with microlensing and stellar rotation curve observations \cite{Iocco:2011jz}.  The model is free from possible other observational bounds especially from the continuum gamma-ray flux and also antiprotons. We also discussed possible production mechanisms for the dark matter from the inflation or a heavy particle decay but other possibilities are still open for  studies in the future.

\acknowledgements{SC is supported by Basic Science Research
Program through the National Research Foundation of Korea funded
by the Ministry of Education, Science and Technology
(2011-0010294) and (2011-0029758).}

\begin{appendix}
\section{Kinetic mixing between $\rm U(1)_{EM}$ and massless $\rm U(1)_H$}\label{Appendix}
\label{sec:appendix}

Consider two Abelian gauge groups $\rm U(1)_{EM}$ and
$\rm U(1)_H$.\footnote{For the case of massive $\rm U(1)_H$, see
Ref.~\cite{Chun:2010ve}.} The kinetic mixing between $\rm U(1)_{EM}$
and $\rm U(1)_H$ is parameterized as
\begin{eqnarray}\label{U(1)emU(1)ex}
{\cal L} = -\frac{1}{4}\, \hat{F}_{\mu\nu}\hat{F}^{\mu\nu}
- \frac{1}{4}\, \hat{F}_{H \mu\nu}\hat{F}_H^{\mu\nu}
- \frac{\xi}{2}\, \hat{F}_{\mu\nu}\hat{F}_H^{\mu\nu}\,,
\end{eqnarray}
where $\hat{A}_\mu (\hat{A}^H_{\mu})$ is the $U(1)_{EM} (U(1)_H)$
gauge boson and its field strength tensor is $\hat{F}^{\mu\nu}
(\hat{F}^{\mu\nu}_H)$. The kinetic mixing is parameterized by
$\xi$, which is generically allowed by the gauge invariance and the
Lorentz symmetry. In the low energy effective theory, the kinetic
mixing parameter $\xi$ is considered to be an arbitrary parameter.
An ultraviolet theory is expected to generate the kinetic mixing
parameter $\xi$ \cite{KineticMixing}.

We are only interested in a small kinetic mixing, $\xi < 1$. Thus,
for convenience we can set $\xi \equiv \sin\epsilon$. The
kinetic terms for the photon and hidden photon are diagonalized by the
following transformation:
\begin{eqnarray}
\left(
  \begin{array}{c}
    A'_\mu \\
    A'^{H}_\mu \\
  \end{array}
\right) = \left(
            \begin{array}{cc}
              \cos\frac{\epsilon}{2} & \sin\frac{\epsilon}{2} \\
              \sin\frac{\epsilon}{2} & \cos\frac{\epsilon}{2} \\
            \end{array}
          \right) \left(
                    \begin{array}{c}
                      \hat{A}_\mu \\
                      \hat{A}^H_\mu \\
                    \end{array}
                  \right)\,.
\end{eqnarray}
In this transformed basis, the Lagrangian is given by
\begin{eqnarray}
{\cal L} =
-\frac{1}{4}\, F'_{\mu\nu}F'^{\mu\nu} - \frac{1}{4}\, F'^{H}_{\mu\nu} F'^{H \mu\nu}\,,
\end{eqnarray}
where $F'_{\mu\nu}$ and $F'^{H}_{\mu\nu}$ are the field strength
tensors corresponding to $A'_\mu$ and $A'^{H}_\mu$, respectively.
Since two gauge bosons are massless, we still have an $SO(2)$
symmetry:
\begin{eqnarray}
\left(
  \begin{array}{c}
    A_\mu \\
    A^H_\mu \\
  \end{array}
\right) = \left(
            \begin{array}{cc}
              \cos\eta & \sin\eta \\
              -\sin\eta & \cos\eta \\
            \end{array}
          \right) \left(
                    \begin{array}{c}
                      A'_\mu \\
                      A'^{H}_\mu \\
                    \end{array}
                  \right)\,.
\end{eqnarray}
If we choose the mixing angle as $\eta = -\epsilon/2$, the final
relation between the bases $(A_\mu, A^H_\mu)$ and $(\hat{A}_\mu,
\hat{A}^H_\mu)$ is
\begin{eqnarray}
\left(
  \begin{array}{c}
    A_\mu \\
    A^H_\mu \\
  \end{array}
\right) = \left(
            \begin{array}{cc}
              \cos\epsilon & 0 \\
              \sin\epsilon & 1 \\
            \end{array}
          \right) \left(
                    \begin{array}{c}
                      \hat{A}_\mu \\
                      \hat{A}^H_\mu \\
                    \end{array}
                  \right)\,.
\end{eqnarray}
By this diagonalization procedure of the kinetic terms, we finally obtain
\begin{eqnarray}
{\cal L} =
-\frac{1}{4}\, F_{\mu\nu}F^{\mu\nu} - \frac{1}{4}\, F^H_{\mu\nu} F^{H \mu\nu}\,,
\end{eqnarray}
where the new field strength tensors are  $F_{\mu\nu}$ and
$F_{\mu\nu}^H$. The SM photon corresponds to $A_\mu$ and the
hidden photon to $A^H_\mu$.

Let us take the following simple interaction Lagrangian of a SM
fermion $f$ with the photon in the original basis as
\begin{eqnarray}
{\cal L}_f = \bar{f} \left( \hat{e} Q_f \gamma^\mu \right) f
\hat{A}_\mu\,. \label{SM-interaction}
\end{eqnarray}
Note that in this basis no direct interaction exists between the
SM fermion and the hidden sector gauge boson $\hat{A}_H$. If there
exists a hidden sector fermion $\psi$ with the $\rm U(1)_H$ charge
$Q^H_{\psi}$, its interaction with the hidden sector gauge boson
is simply represented by
\begin{eqnarray}
{\cal L}_\psi = \bar{\psi} \left( \hat{g}_{H} Q^H_{\psi} \gamma^\mu
\right) \psi \hat{A}^H_{\mu}\,. \label{hidden-interaction}
\end{eqnarray}
In this case, there is also no direct interaction between
the hidden fermion and the SM photon $\hat{A}$.

We can recast the Lagrangian (\ref{SM-interaction}) in the
transformed basis $(A_\mu, A^H_\mu)$,
\begin{eqnarray}
{\cal L}_f = \bar{f} \left( \frac{\hat{e}}{\cos\epsilon} Q_f
\gamma^\mu \right) \psi A_\mu\,.
\end{eqnarray}
Here, one notices that the SM fermion has a coupling only to the
visible sector gauge boson $A$ even after changing the basis of
the gauge bosons. However, the coupling constant $\hat{e}$ is
modified to $\hat{e} / \cos\epsilon$, and so the physical visible
sector coupling $e$ is just defined as $e \equiv \hat{e} /
\cos\epsilon$. Similarly, we can derive the following interactions
for $\chi$,
\begin{eqnarray} {\cal L}_\psi = \bar{\psi} \gamma^\mu
\left(\hat{g}_H Q^H_{\psi} A^H_\mu - \hat{g}_H \tan\epsilon
Q_{\psi} A_\mu \right) \psi\,. \label{hidden-shift}
\end{eqnarray}
In this basis, the hidden sector matter field $\psi$ now can
couple to the SM photon $A$ with the coupling $-\hat{g}_H
Q^H_{\psi} \tan\epsilon$. Consequently, we can interpret the
hidden particle $\psi$ as a particle with a EM charge $Q_\psi
\equiv (-\hat{g}_H Q^H_{\psi} \tan\epsilon) / e$. In addition, we
can set the physical hidden sector coupling $g_H$ as $g_H \equiv
\hat{g}_H$.

\end{appendix}


\begin{thebibliography}{99}

\bibitem{Jungman:1995df}
  G.~Jungman, M.~Kamionkowski and K.~Griest,
  Phys.\ Rept.\  {\bf 267} (1996) 195  [hep-ph/9506380];
  G.~Bertone, D.~Hooper and J.~Silk,
  Phys.\ Rept.\  {\bf 405}, 279 (2005)  [hep-ph/0404175].

\bibitem{Scalar}
  V.~Silveira and A.~Zee,
  Phys.\ Lett.\ B {\bf 161}, 136 (1985);
  J.~McDonald,
  Phys.\ Rev.\ D {\bf 50}, 3637 (1994)  [hep-ph/0702143 [HEP-PH]];
  C.~P.~Burgess, M.~Pospelov and T.~ter Veldhuis,
  Nucl.\ Phys.\ B {\bf 619}, 709 (2001)  [hep-ph/0011335].

\bibitem{Fermion}
  Y.~G.~Kim and K.~Y.~Lee,
  Phys.\ Rev.\ D {\bf 75}, 115012 (2007)  [hep-ph/0611069];
  Y.~G.~Kim, K.~Y.~Lee and S.~Shin,
  JHEP {\bf 0805}, 100 (2008)  [arXiv:0803.2932 [hep-ph]].

\bibitem{millicharged}
  K.~Cheung and T.~-C.~Yuan,
  JHEP {\bf 0703}, 120 (2007)  [hep-ph/0701107];
  D.~Feldman, Z.~Liu and P.~Nath,
  Phys.\ Rev.\ D {\bf 75}, 115001 (2007)  [hep-ph/0702123 [HEP-PH]].

\bibitem{511millicharged}
  J.~-H.~Huh, J.~E.~Kim, J.~-C.~Park and S.~C.~Park,
  Phys.\ Rev.\ D {\bf 77}, 123503 (2008)  [arXiv:0711.3528 [astro-ph]].

\bibitem{Vector}
  M.~Pospelov, A.~Ritz and M.~B.~Voloshin,
  Phys.\ Rev.\ D {\bf 78}, 115012 (2008)  [arXiv:0807.3279
  [hep-ph]].

\bibitem{Chen}
  C.~-R.~Chen, F.~Takahashi and T.~T.~Yanagida,
  Phys.\ Lett.\ B {\bf 671}, 71 (2009)  [arXiv:0809.0792 [hep-ph]];
  C.~-R.~Chen, F.~Takahashi and T.~T.~Yanagida,
  Phys.\ Lett.\ B {\bf 673}, 255 (2009)  [arXiv:0811.0477 [hep-ph]];
  C.~-R.~Chen, M.~M.~Nojiri, F.~Takahashi and T.~T.~Yanagida,
  Prog.\ Theor.\ Phys.\  {\bf 122}, 553 (2009)  [arXiv:0811.3357 [astro-ph]].

\bibitem{Fermi}
  W.~B.~Atwood {\it et al.}  [LAT Collaboration],
  Astrophys.\ J.\  {\bf 697}, 1071 (2009)  [arXiv:0902.1089 [astro-ph.IM]].

\bibitem{Bringmann:2012vr}
  T.~Bringmann, X.~Huang, A.~Ibarra, S.~Vogl and C.~Weniger,
  arXiv:1203.1312 [hep-ph].

\bibitem{Weniger:2012tx}
  C.~Weniger,
  arXiv:1204.2797 [hep-ph].

\bibitem{Su:2012ft}
  M.~Su and D.~P.~Finkbeiner,
  arXiv:1206.1616 [astro-ph.HE].

\bibitem{Tempel:2012ey}
  E.~Tempel, A.~Hektor and M.~Raidal,
  arXiv:1205.1045 [hep-ph].

\bibitem{Tempel2}
   A.~Hektor, M.~Raidal and E.~Tempel,
  arXiv:1207.4466 [astro-ph.HE].


\bibitem{Ackermann:2012qk}
  M.~Ackermann {\it et al.}  [LAT Collaboration],
  arXiv:1205.2739 [astro-ph.HE].


\bibitem{GeringerSameth:2012sr}
  A.~Geringer-Sameth and S.~M.~Koushiappas,
  arXiv:1206.0796 [astro-ph.HE].


\bibitem{Boyarsky:2012ca}
  A.~Boyarsky, D.~Malyshev and O.~Ruchayskiy,
  arXiv:1205.4700 [astro-ph.HE].

\bibitem{FermiBubble}
  S.~Profumo and T.~Linden,
  arXiv:1204.6047 [astro-ph.HE];
  M.~Su, T.~R.~Slatyer and D.~P.~Finkbeiner,
  Astrophys.\ J.\  {\bf 724}, 1044 (2010)  [arXiv:1005.5480
  [astro-ph.HE]].








\bibitem{Ibarra:2012dw}
  A.~Ibarra, S.~Lopez Gehler and M.~Pato,
  arXiv:1205.0007 [hep-ph].

\bibitem{Dudas:2012pb}
  E.~Dudas, Y.~Mambrini, S.~Pokorski and A.~Romagnoni,
  arXiv:1205.1520 [hep-ph].

\bibitem{Cline:2012nw}
  J.~M.~Cline,
  arXiv:1205.2688 [hep-ph].

\bibitem{KYChoi}
  K.~Y.~Choi and O. Seto,
  arXiv:1205.3276 [hep-ph].

\bibitem{Lee:2012bq}
  H.~M.~Lee, M.~Park and W.~-I.~Park,
  arXiv:1205.4675 [hep-ph].

\bibitem{Rajaraman:2012db}
  A.~Rajaraman, T.~M.~P.~Tait and D.~Whiteson,
  arXiv:1205.4723 [hep-ph].

\bibitem{Acharya:2012dz}
  B.~S.~Acharya, G.~Kane, P.~Kumar, R.~Lu and B.~Zheng,
  arXiv:1205.5789 [hep-ph].

\bibitem{Buckley:2012ws}
  M.~R.~Buckley and D.~Hooper,
  arXiv:1205.6811 [hep-ph].

\bibitem{Chu:2012qy}
  X.~Chu, T.~Hambye, T.~Scarna and M.~H.~G.~Tytgat,
  arXiv:1206.2279 [hep-ph].

\bibitem{Das:2012ys}
  D.~Das, U.~Ellwanger and P.~Mitropoulos,
  arXiv:1206.2639 [hep-ph].

\bibitem{Kang:2012bq}
  Z.~Kang, T.~Li, J.~Li and Y.~Liu,
  arXiv:1206.2863 [hep-ph].

\bibitem{Oda:2012fy}
  I.~Oda,
  arXiv:1207.1537 [hep-ph].

\bibitem{Yang:2012ha}
  R.~-Z.~Yang, Q.~Yuan, L.~Feng, Y.~-Z.~Fan and J.~Chang,
  arXiv:1207.1621 [astro-ph.CO].









\bibitem{Kyae:2012vi}
  B.~Kyae and J.~-C.~Park,
  arXiv:1205.4151 [hep-ph].









\bibitem{Buchmuller:2012rc}
  W.~Buchmuller and M.~Garny,
  arXiv:1206.7056 [hep-ph].


\bibitem{Cohen:2012me}
  T.~Cohen, M.~Lisanti, T.~R.~Slatyer and J.~G.~Wacker,
  arXiv:1207.0800 [hep-ph].

\bibitem{Cholis:2012fb}
  I.~Cholis, M.~Tavakoli and P.~Ullio,
  arXiv:1207.1468 [hep-ph].




\bibitem{Einasto}
J. Einasto. Trudy Inst. Astroz. Alma-Ata : 51 (1965) 87.


\bibitem{NFW} 
  J.~F.~Navarro, C.~S.~Frenk and S.~D.~M.~White,
  Astrophys.\ J.\  {\bf 462}, 563 (1996)
  [astro-ph/9508025].




\bibitem{Goldberg:1986nk}
  H.~Goldberg and L.~J.~Hall,
  ``A New Candidate For Dark Matter,''
  Phys.\ Lett.\ B {\bf 174}, 151 (1986).



\bibitem{LandauYang}
  L. D. Landau, Dokl. Akad. Nawk., USSR {\bf 60}, 207 (1948);
  C.~-N.~Yang,
    Phys.\ Rev.\  {\bf 77}, 242 (1950).  




\bibitem{Garny:2010eg}
  M.~Garny, A.~Ibarra, D.~Tran and C.~Weniger,
  JCAP {\bf 1101} (2011) 032  [arXiv:1011.3786 [hep-ph]].



\bibitem{KineticMixing}
  L.~B.~Okun,
  Sov.\ Phys.\ JETP {\bf 56}, 502 (1982)
  [Zh.\ Eksp.\ Teor.\ Fiz.\  {\bf 83}, 892 (1982)];
  B.~Holdom,
  Phys.\ Lett.\  B {\bf 166}, 196 (1986).


\bibitem{Langacker08}
 For a review, see, P.~Langacker,
  ``The Physics of Heavy $Z^\prime$ Gauge Bosons,''
  Rev.\ Mod.\ Phys.\  {\bf 81} (2008) 1199


\bibitem{Chun:2010ve}
  E.~J.~Chun, J.~-C.~Park and S.~Scopel,
  JHEP {\bf 1102}, 100 (2011)  [arXiv:1011.3300 [hep-ph]].


\bibitem{HiddenPhoton}
  J.~Jaeckel and S.~Roy,
  Phys.\ Rev.\ D {\bf 82}, 125020 (2010)  [arXiv:1008.3536 [hep-ph]].



\bibitem{Iocco:2011jz} 
  F.~Iocco, M.~Pato, G.~Bertone and P.~Jetzer,
  JCAP {\bf 1111}, 029 (2011)
  [arXiv:1107.5810 [astro-ph.GA]].


\end{thebibliography}
\end{document}